\documentclass[12pt,preprint]{aastex}
\shorttitle{Masers in Blazars}
\shortauthors{Begelman, Ergun \& Rees}

\begin{document}

\title{Cyclotron Maser Emission from Blazar Jets?}

\author{Mitchell C. Begelman$^{1,4}$, Robert E. Ergun$^{2,4}$, and Martin J. Rees$^3$}
\affil{
$^1$Joint Institute for Laboratory Astrophysics, University of Colorado,\\
Boulder, CO 80309-0440, USA \\ $^2$Laboratory for Atmospheric and Space Physics, University of Colorado,\\
Boulder, CO 80309-0392, USA \\ $^3$Institute of Astronomy, Madingley Road, Cambridge CB3 0HA, UK \\
$^4$Also at Department of Astrophysical and Planetary Sciences, University of Colorado at Boulder 
 \email{mitch@jila.colorado.edu; ree@lasp.colorado.edu; mjr@ast.cam.ac.uk}
}

\begin{abstract}

We consider the production of electron cyclotron maser emission by low-density, highly magnetized plasmas in relativistic jets. The population inversion required to drive cyclotron maser instability could occur in localized, transient sites where hydromagnetic instabilities, shocks, and/or turbulence lead to magnetic mirroring along current-carrying flux tubes. The maser is pumped by the conversion of kinetic and magnetic energy into ${\bf j}\cdot {\bf E}$ work as electrons are accelerated into the maser by the parallel electric field that develops as a result of the mirror.  We estimate the maximum brightness temperatures $T_B$ that can be obtained in a single maser site and in an array of many masers operating simultaneously, under conditions likely to apply in blazar jets.  Synchrotron absorption, by relativistic electrons within the jet, presents the largest obstacle to the escape of the maser radiation, and may render most of it invisible.  However, we argue that a high brightness temperature could be produced in a thin boundary layer outside the synchrotron photosphere, perhaps in the shear layer along the wall of the jet.  Induced Compton scattering provides additional constraints on the maximum brightness temperature of a masing jet.  We suggest that recent observations of diffractive scintillation in the blazar J1819+3845, indicating intrinsic brightness temperatures greater than $10^{14}$ K at 5 GHz, may be explained in terms of cyclotron maser emission. High-$T_B$ maser emission from blazar jets may extend to frequencies as high as $\sim 100$ GHz, with the maximum possible $T_B$ scaling with frequency roughly $\propto \nu^{-1}$. Less massive relativistic jet sources, such as microquasars, are even better candidates for producing cyclotron maser emission, primarily in the infrared and optical bands. 

\end{abstract}
 
\keywords{acceleration of particles---masers---radiation mechanisms: nonthermal---galaxies: jets---radio continuum: galaxies}

\section{Introduction}

There has been a longstanding debate over whether some compact extragalactic radio sources have intrinsic brightness temperatures that exceed the maximum value consistent with incoherent synchrotron radiation.  The maximum brightness temperature for steady emission, $\sim 10^{12}$ K, is set by the ``inverse Compton catastrophe" (Kellermann \& Pauliny-Toth 1969), although attainment of this limit requires very large deviations from equipartition between magnetic and particle energy and, therefore, unacceptably high energy requirements.  A more plausible upper limit, $\sim 10^{11}$ K, is obtained for self-absorbed synchrotron sources close to equipartition; most resolved compact sources fall near or below this limit (Readhead 1994).

However, the observational limits on brightness temperature are not as clear-cut for bright, unresolved radio sources that exhibit rapid ``intraday variability" (IDV) at GHz frequencies.  If simple light-travel time arguments are used to infer an angular size from the variability timescale, then some of these sources must have brightness temperatures as high as $T_B \sim 10^{21}$ K.  The actual angular size of the source can be a factor $\sim \Gamma$ larger than the naive estimate if the radiating material (or, rather, its emission pattern) approaches us with a bulk Lorentz factor $\Gamma \gg 1$.  The solid angle is thus a factor $\sim \Gamma^2$ larger, while the intrinsic brightness temperature of the beam is boosted by a factor $\Gamma$ relative to the comoving frame of the source.  Thus, the apparent $T_B$ could be boosted by a factor $\sim \Gamma^3$ over its value in the comoving frame, and values of $\Gamma \sim 10^3$ could explain most extreme instances of IDV.  However, such large bulk Lorentz factors also imply very low synchrotron radiative efficiencies, leading to implausibly large jet energy fluxes (Begelman, Rees \& Sikora 1994).

The need for high intrinsic brightness temperatures appeared to become less acute when it was confirmed that IDV results mainly from refractive interstellar scintillation (Kedziora-Chudczer et al. 1997; Dennett-Thorpe \& de Bruyn 2000, 2002). These observations do not require intrinsic brightness temperatures larger than $\sim 10^{13}$ K, which are easily explained by incoherent synchrotron radiation boosted by a modest Lorentz factor $\la 30$.  However, space VLBI has failed to resolve these and many other compact sources (Tingay et al. 2001), leaving open the possibility that the brightness temperatures are much higher. Moreover, recent observational evidence for {\it diffractive} scintillation from the blazar J1819+3845 (Macquart \& de Bruyn 2004) yields an upper limit for the angular size of $< 10 \ \mu$as, implying intrinsic brightness temperatures in excess of $ 10^{14}$ K.  To explain such high intrinsic $T_B$ with an incoherent synchrotron model would require $\Gamma \ga 10^3$, once again leading to implausible energy requirements. 
 
Brightness temperatures that exceed the synchrotron limit, after relativistic effects are taken into account, could indicate that a coherent radiation mechanism is at work. There have been several attempts to apply such mechanisms to the conditions in relativistic jets from active galactic nuclei (e.g., Baker et al. 1988; Sol, Pelletier \& Asseo 1989; Krishnan \& Wiita 1990; Lesch \& Pohl 1992; Benford 1992; Benford \& Tzach 2000). All of these models rely on the development of strong Langmuir turbulence, and implicitly assume that the electron plasma frequency $\nu_p$ exceeds the electron cyclotron frequency $\nu_c$.  But simple models of relativistic jets in blazars suggest that just the opposite is the case (see \S 3.2).  In this spirit, we propose that high-$T_B$ radiation could result from electron cyclotron maser instability, which operates efficiently when $\nu_c \gg \nu_p$ and which has been applied successfully to explain auroral kilometric radiation, Jovian decametric radiation, and possibly certain stellar flares (Melrose 1999).  

In \S 2 we describe how small-scale, current-carrying magnetic mirrors, which might result from hydromagnetic instabilities, shocks and/or turbulence, set up ideal conditions for electron cyclotron maser emission. As current propagates into a mirror, the motion of the current-carrying electrons along the flux tube is inhibited.  To maintain the current through the mirror, a parallel electric field is established.  The motion of accelerated electrons along converging flux tubes automatically creates the population inversion required for a cyclotron maser to operate. We then show that conditions thought to exist in blazar jets are favorable for this process to operate (\S 3).  We estimate the fraction of the jet that needs to be filled with masers to produce a given observed brightness temperature, and show that even modest relativistic beaming (with a bulk Lorentz factor $\la 10$) greatly increases the efficiency of the mechanism.  Of the several processes that can inhibit the escape of high-$T_b$ maser emission, synchrotron absorption is potentially the most serious.  In  \S 3.2 we show that it is likely to prevent most of the maser radiation from reaching us, but argue that high brightness temperatures can still be produced outside the synchrotron photosphere, in a thin boundary layer along the interface between the jet and the ambient medium.  Induced Compton scattering provides an additional constraint on the maximum brightness temperature of a masing jet, while moderate relativistic beaming greatly facilitates the escape of the radiation.  Finally (\S 4), we summarize our main results, generalize the model to the production of maser emission at frequencies outside the GHz range, and discuss how the maser emission could scale with the mass of the system.   We find that Galactic microquasars are particularly promising sources of cyclotron maser radiation, at frequencies that could range upward to the infrared or optical range. 
 
\section{Electron Cyclotron Maser Emission}

\subsection{General conditions for cyclotron masers}

Cyclotron maser emission is thought to be produced by mildly relativistic electrons at frequencies close to the cyclotron fundamental or its low harmonics.  Relativistic corrections to the gyrofrequency are essential to the operation of the maser instability (Wu \& Lee 1979, Holman, Eichler \& Kundu 1980, Melrose \& Dulk 1982), which loses efficiency if the electron energy is too low ($\la 10$s of keV).  Masers can operate at relativistic energies (Louarn, Le Queau \& Roux 1986), but it is possible that the effect weakens if the electron are too relativistic, since the resonances can become too broad.  Efficient cyclotron maser emission also requires a highly magnetized plasma, in the sense that the electron plasma frequency, $\nu_p \propto n_e^{1/2}$, is much smaller than the cyclotron frequency, $\nu_c$.  Emission occurs at close to the cyclotron frequency if the condition $\nu_p \ll \nu_c$ is met.  
 
Several schemes have been proposed to obtain the kind of population inversion required for a cyclotron maser.  The most common approach, called the ``loss-cone" mechanism, is to suppose that a distribution of electrons with an initially isotropic pitch-angle distribution propagates down converging magnetic field lines toward an absorbing boundary, such as a stellar or planetary atmosphere (Holman et al. 1980, Melrose \& Dulk 1982, Willes \& Wu 2004).  The particles with sufficiently low pitch angles reach the boundary and are absorbed, while those with pitch angles above a certain threshold are reflected by the magnetic mirror.  Loss-cone masers have the advantage that they take an initially isotropic electron distribution function and naturally turn it into an inverted distribution through the selective loss of particles in a fixed magnetic field geometry.  However, loss-cone inversions are readily quenched by pitch-angle diffusion, caused by the radiation process.  Since the loss cone can be filled in without strong radiative losses, loss-cone masers tend to be inefficient (Melrose 2002). 
  
Several authors have recently focused on the ``shell" or ``horseshoe" mechanism, in which the electron distribution is driven away from isotropy by a coherent acceleration mechanism (Pritchett 1984, Winglee \& Pritchett 1986, Louarn et al. 1990, Delory et al. 1998, Pritchett et al. 1999, Bingham \& Cairns 2000, Bingham et al. 2003).  Since the low-energy electrons are depleted by the acceleration process, the distribution function resembles a shell or horseshoe on a $v_\parallel$ vs.~$v_\perp$ plot (see Fig.~1). This mechanism is more robust than the loss-cone mechanism (Pritchett et al. 1999), because filling in the shell (and thus quenching the inversion) cannot occur without strong energy losses.  The maser instability growth rates tend to be higher and the inverted distribution can be pumped vigorously and continuously by a parallel electric field driving a current into a mirroring field geometry.  Ergun et al. (2000) found strong evidence from in situ observations that auroral kilometric radiation results from a shell maser, rather than a loss-cone maser.  No absorbing boundary is needed for a shell maser, which is an advantage for our application since such a boundary is unlikely to exist in the low-density regions of a relativistic jet. 
 
\subsection{Cyclotron maser emission from current-carrying magnetic mirrors} 

We suppose that a blazar jet contains strongly magnetized, highly turbulent plasma. The turbulence could be driven by boundary conditions at the base of the jet (e.g., flux loops erupting from an accretion disk), hydromagnetic instabilities, strong pressure gradients, or velocity shear internal to the jet or near the jet boundary.  The result is a highly tangled magnetic field topology containing strong, mostly field-aligned currents, current sheets, and flux ropes such as seen in the interstellar medium (e.g., Spangler 1999) and in active regions in the solar corona (e. g., Dmitruk, G\'omez \& Deluca 1998). These currents inevitably localize and dissipate in relatively confined regions (Parker 1988). As a result, large contrasts in magnetic field strength and direction are inescapable. A change in magnetic field strength along a flux rope sets up a magnetic mirror.   
Field-aligned currents in magnetic mirrors are expected in regions of oblique shocks (Bingham et al. 2003), magnetic field reversals associated with current sheets, strong pressure gradients, boundaries and shear layers, magnetic reconnection, or instabilities (kink or sausage) related to quasi-force-free flux ropes.  Such magnetic field topologies, for example, are often observed in the solar wind. Magnetic field measurements in the solar wind during active periods (e.g. Tsurutani et al. 1995; Lepping et al. 1997) display order of magnitude variations in $|{\bf B}|$ and rapid (or small-scale) changes in magnetic field direction. The large variations in the magnetic fields are often associated with erupting flux ropes (e.g., Lepping et al. 1997) or shocks in corotating interaction regions (Tsurutani et al. 1995). It is a basic premise of this article that current-carrying mirrors are fundamental to strongly turbulent plasmas but the dominant mechanism that creates the field-aligned currents and changes in $|{\bf B}|$, while interesting and important, is for future debate.

Below, we demonstrate that a field-aligned current in a magnetic mirror creates ideal conditions for a shell maser. The current-carrying magnetic mirror sets up a parallel electric field and causes the accelerated electron distribution to evolve into a shell in velocity-space which has been shown to produce powerful maser emissions. In a strong magnetic mirror (mirror ratio $R \ga 10$), such as in the stellar or planetary cases, the parallel potential is large resulting in strong electron acceleration. Furthermore, the electron distribution can fully evolve into an unstable shell distribution resulting in a powerful maser. A weak magnetic mirror ($R \la 2$), however, leads to weaker electron energization and the electron distribution does not fully evolve into a shell, so the maser is far less powerful. In a strongly turbulent plasma, the current within the flux ropes or current sheets will encounter a wide range of magnetic mirrors, the vast majority of which will be weak.  We show, however, that a relatively sparse packing of moderate magnetic mirrors ($R \sim 5$) can produce an extremely bright ($T_B \sim 10^{15}$ K) source.   To quantify the problem, we focus on the simple case of a slender, quasi-force-free flux rope characterized by a transverse (toroidal) magnetic field $B_0$ and a radius $r_0$ that has a transverse (toroidal) mirroring field $B_m$ with $R \equiv B_m/B_0 \sim 5$. A cartoon depicting the flux rope is shown in Figure 1. If the mirror has a characteristic radius $r_m$, the longitudinal current density required to maintain this external field is
\begin{equation}
\label{jpar}
j_0 \sim {c\over 4\pi}{B_0 \over r_0} \ \ {\rm outside\ the\ mirror; \ and}\  j_m \sim {c\over 4\pi}{B_m \over r_m} \ \ {\rm inside\ the\ mirror}. 
\end{equation}

In a mildly relativistic (non-positron) plasma, electrons carry the majority of the field-aligned currents. As the flux rope evolves, the mirror force restricts the available phase space in the electron distribution outside of the mirror region that can carry currents through the region of high magnetic field (Fridman and Lemaire 1980). As in the case of Earth (e.g., Weimer et al. 1987, Lyons 1980, Elphic et al. 1998) and Jupiter (e.g. Su et al., 2003 and references therein), a parallel electric field will develop if the mirror ratio is such that   
\begin{equation}
\label{jtherm}
j_0 = {c\over 4\pi}{B_0 \over r_0} > {n_0 \beta_e e c \over R}, 
\end{equation}
where $n_0$ is the electron density in the flux rope and $\beta_e c \approx (kT_e/m_e)^{1/2}$ is the electron thermal speed.  Since strong maser activity requires mildly relativistic electrons, we will assume electron temperatures $T_e \ga 100$ keV.  Such ambient temperatures could be maintained by dissipation at the current sheets, given that the radiative cooling time scales are typically longer than flow times under blazar conditions.  If $R< (m_i/m_e)^{1/2}(T_e/T_i)^{1/2}$ (where subscript $i$ refers to the ions), the electron drift speed exceeds the ion thermal speed. Such currents are ubiquitous in the Earth's auroral zone and magnetosphere (e.g., McFadden, Carlson \& Ergun 1999). Electron drift speeds routinely are observed to exceed ion thermal speeds and approach, but rarely exceed, the electron thermal speed. In observed regions of electron cyclotron maser activity, these high drift speeds excite a number of plasma modes including acoustic turbulence, ion cyclotron waves, lower hybrid waves, and quasi-electrostatic whistler emissions (see Kindel \& Kennel 1970 for a review). These emissions, however, do not offer sufficient resistivity to disrupt the electron current nor do they hinder the electron cyclotron maser process.   
 
The electric potential $\Phi$ along the magnetic field required to maintain the current through the pinch can be expressed as 
\begin{equation}
\label{jknight}
j_m = Rj_0 = {\Phi e^2 n_0 \over (2\pi m_e k T_e)^{1/2} } \ \  {\rm with} \ \ 1 \leq {e\Phi \over k T_e } \leq R  
\end{equation}
(Knight 1973, Fridman and Lemaire 1980) which, when combined with eq.~(\ref{jpar}), becomes
\begin{equation}
\label{jknight2}
{B_m \over 4\pi e n_0 r_m} = {e \Phi \over (2\pi)^{1/2} \beta_e m_e c^2} = \left({e\Phi \over 4\pi k T_e }\right)^{1/2} \left({2 e \Phi \over m_e c^2}\right)^{1/2} \ .
\end{equation}
The parallel electric field increases the current density to the required level by accelerating electrons along the magnetic field lines.  The right hand side of eq.~(\ref{jknight2}) is broken into two factors. The second factor is simply the electron drift velocity from acceleration by the parallel potential. The leading factor is a correction due to the mirroring geometry which slows the electrons as they move into regions of high magnetic field.  

The above equations are valid in the subrelativistic limit. The growth rate of the maser instability increases steeply with electron energy up to relativistic energies, and levels off or declines at higher energies. If the initial electron energies are already mildly relativistic (our reason for assuming $T_e \ga 100$ keV), then acceleration up to electron Lorentz factors $\gamma \ga 2$ requires modest mirror ratios, $R \sim 5$. The maser instability is so powerful that whenever it operates, electrons entering the pinch are unlikely to exceed mildly relativistic energies and we will use the nonrelativistic formulae throughout. We will therefore assume, for purposes of quantitative estimates below, that 
$e \Phi \sim  5 kT_e \sim m_e c^2$ and $B_m / (4\pi e n_0 r_m) = j_m / (e c n_0 ) \sim 1$.  

Because the increase in energy is the same for all electrons, the Maxwellian velocity distribution is not preserved in this acceleration process. In addition to the characteristic energy of the distribution shifting to higher values, the low-energy electron population is depleted, leading to a situation in which $\partial f({\bf v})/ \partial v_\parallel >0$. Although this represents a population inversion, it is not one that favors cyclotron maser instability.  In order to obtain a cyclotron maser a second step is required in the evolution of the electron distribution function --- conversion of the inversion in $v_\parallel$ to one in $v_\perp$ (i.e., $\partial f({\bf v})/ \partial v_\perp >0$).  This occurs naturally if the accelerated electron current propagates into a region of sufficiently increasing magnetic field strength: a magnetic mirror.
 
Figure 1 illustrates the population inversion process. The parallel electric field forms on the low-potential side of the pinch region as on Earth and Jupiter. The unaccelerated electron distribution has a small loss cone which is weakly unstable to the maser (Fig.~1b). The accelerated electron distribution rapidly evolves in pitch-angle, conserving energy $(m_e^2 c^4 + p^2 c^2)^{1/2}$ and magnetic moment  $ \mu = p_\perp^2/(me|{\bf B}|)$, where ${\bf p}$ is the momentum. Near the highest field region of the mirror, the distribution resembles a horseshoe in 2-D or a hollow shell with a loss cone in 3-D (Fig.~1c). The resonance condition of the maser (dashed line in Fig.~1c) is for purely X-mode emission with a frequency $\nu = \nu_c/\gamma$. The shell maser draws its energy from the entire inverted electron population and can therefore convert a substantial fraction of the ${\bf j}\cdot {\bf E}$ power into electromagnetic waves. 
 
Both the acceleration and mirroring are likely to occur in close proximity or co-spatially.  Once a strong $v_\perp-$inversion occurs, the stimulated emission of cyclotron waves runs away on a very short timescale and quickly saps the available free energy, filling in the low-energy portion of the electron distribution and essentially thermalizing the electrons as they pass through the mirror region.  However, this does not quench the maser: the shell maser process is continuous (Ergun et al. 2000b) with the maser output proportional to the supply of ${\bf j}\cdot {\bf E}$ work associated with particle acceleration through the mirror.  

\subsection{Maser emissivity}

The emission from a single cyclotron maser is highly anisotropic.  In the case of a shell maser it is strongly concentrated in the plane normal to the local magnetic field. While the spectrum peaks at just below the cyclotron frequency, the bandwidth $\Delta \nu/ \nu \sim (\gamma - 1)/\gamma$ (where $\gamma$ is the random electron Lorentz factor) is as large as $1/2$ if the radiating electrons are moderately relativistic ($\gamma \sim 2$).  In practice, the detailed emission properties of any one maser site is unimportant, because the observed radiation will be the incoherent sum of emission from an extremely large number of masers.  The angular distribution of radiation will be largely isotropized by the randomness of the magnetic field directions and the overall bandwidth of coherent radiation will be determined primarily by the range of magnetic field strengths. A randomized distribution of magnetic field directions would also largely wash out the strong circular polarization associated with a single cyclotron maser, although any bias in the mean field properties (e.g., a preferred direction relative to the line of sight) should show up in the polarization. In the following, therefore, we will estimate the properties of a typical maser assuming that it radiates isotropically. 

The power of a single maser is a fraction $\xi$ of the power dissipated in the mirror,
\begin{equation}
\label{power1}
P_m  =  \xi \int {\bf j}\cdot {\bf E}\  dV  \sim \pi \xi j_m \Phi r_m^2 ,
\end{equation}
where we have assumed that the acceleration volume is cylindrical with radius $r_m$.  Based on our discussion above, we estimate $\Phi \sim m_e c^2/ e$ and adopt $j_m \sim c B_m / (4\pi r_m) \sim  n_0 e c $. The efficiency of a shell maser, $\xi$, is limited thermodynamically to $\xi_{\rm max} \approx 0.6 (e\Phi - kT_e)/e\Phi$, but this thermodynamic limit is never reached in practice.  The maser efficiency can be reduced by competing modes that draw their energy from the unstable shell distribution. However, almost all of the competing modes are at frequencies below the electron plasma frequency, so the maser condition $\nu_p \ll \nu_c$ (Melrose \& Dulk 1982) ensures that growth rates of competing modes are small.  In the case of the Earth's aurora, with $\nu_p / \nu_c \sim 10^{-2}$, satellite measurements indicate that the maser efficiency can be as high as a $\sim 1$ per cent (Gurnett 1974). In the case of blazar jets, $\nu_p / \nu_c$ could be even smaller (see \S 3.2) and the efficiency could be as large as $\ga 10 \% $. We will adopt the normalization $\xi_{-2} = \xi / 0.01$.

Expressing $n_0$ in units of cm$^{-3}$ and setting $B_2 = B_m / 10^2$ G, we obtain a pinch radius of
\begin{equation}
\label{pinchrad}
r_m  \sim  1.7\times 10^{10} B_2 n_0^{-1} \ {\rm cm} 
\end{equation}
and a power per maser of
\begin{equation}
\label{power}
P_m  \sim  \pi \xi n_0 m_e c^3 r_m^2 \sim 2.1 \times 10^{23} \xi_{-2} B_2^2 n_0^{-1}  \ {\rm erg \ s}^{-1}.
\end{equation}
Although our immediate aim is to explain masers observed at frequencies of a few GHz, we chose a fiducial pinch magnetic field strength of 100 G, corresponding to a cyclotron frequency $\sim 0.3$ GHz, because the maser emission from a blazar jet is expected to be Doppler boosted by a factor $\sim 10$. If the maser is modeled as a sphere of radius $r_m$, the mean intensity is
\begin{equation}
\label{intensity}
I_\nu  \sim  {P_m \over 16 \pi^2 \nu_c r^2_m } \sim 1.7. \times 10^{-8} \xi_{-2} B_2^{-1} n_0  \  {\rm erg \ cm}^{-2} \ {\rm s}^{-1} \ {\rm Hz}^{-1} \ {\rm sr}^{-1} .
\end{equation}
The associated brightness temperature is
\begin{equation}
\label{tb}
T_B  =  {c^2 \over 2 \nu_c^2 k} I_\nu \sim  6.9\times 10^{11}\xi_{-2} B_2^{-3} n_0 \ {\rm K}.
\end{equation}
The observed intrinsic brightness temperature is an average over the unresolved source, given by
\begin{equation}
\label{tbmean}
\langle T_B  \rangle_{\rm obs} =  C_m \Gamma T_B ,
\end{equation}
where $\Gamma$ is the bulk Lorentz factor of the masers toward the observer (more precisely, the Doppler factor) and $C_m$ is the ``covering factor" of masers.  Physically, this is the fraction of the solid angle of the source subtended by masers if $C_m < 1$ and the mean number of masers along a line of sight if $C_m > 1$. 

\section{Application to blazar jets}

\subsection{Relation to jet properties}

Observations of diffractive scintillation of the blazar J1819+3845 (Macquart \& de Bruyn 2004) appear to require intrinsic brightness temperatures $\langle T_B  \rangle_{\rm obs} > 10^{14}$ K at frequencies of a few GHz.  This estimate comes from an upper limit on the size of the scintillating component $\la 10 \ \mu$as; the actual brightness temperature could be far higher. 

We assume that this emission comes from cyclotron masers in a jet moving toward us with a bulk Lorentz factor $\Gamma = 10 \Gamma_1$. The observed frequency of the maser emission is therefore $\nu_{\rm obs}\equiv \nu_{\rm GHz}$ GHz $\sim \Gamma \nu_c \sim 2.8\Gamma_1 B_2$ GHz.   Because the maser emission is produced in magnetic mirrors with $R\sim 5$, the maser frequency can be much larger than the (Doppler-boosted) cyclotron frequency associated with the mean magnetic field strength $B_0$ in the flux rope.  For the purpose of our illustrative numerical examples below, we identify the latter with the mean magnetic field in the jet.  However, one should bear in mind that some maser emission could come from mirrors associated with current sheets where $B_0$ is lower than the mean.  The density in the mirror, which we have taken to be $n_0$, is probably of the same order as the mean jet density. Using the relations given in \S 2.3, we can relate the properties of maser emission to the mean conditions in the jet. 

It is instructive to express the local conditions, $n_0(r)$ and $B_0(r)$, in terms of the energetics of the blazar jet.  We assume that the jet subtends a solid angle $\Omega = 0.1 \Omega_{-1}$ sr as it propagates away from the black hole and parameterize the distance from the black hole $r$ in terms of the gravitational radius of the black hole, $r = x GM/c^2 = 1.5 \times 10^{13} M_8 x$ cm, where $M_8$ is the black hole mass in units of $10^8 M_\odot$. The total jet power $L_j$ can be divided into two parts. The kinetic energy flux is $L_{KE} = \eta n_0 m_i c^3 \Gamma^2 r^2 \Omega \equiv L_j/(1+\sigma)$. Here, $\sigma$ is the magnetization parameter and $\eta$ equals $1/ m_i$ times the mean mass per electron [$= (m_e/m_i) + (n_i/n_0)$, where $n_i$ is the ion density].  The parameter $\eta$ ranges from 1 for normal ion-electron plasma to $m_e/m_i$ if pairs outnumber ions by more than 2000-to-one. The electromagnetic (Poynting) flux is $L_{EM} = (B_0^2/4\pi) c \Gamma^2 r^2 \Omega \equiv \sigma L_j/(1 + \sigma)$.   Letting $L_{45} \equiv L_j/ 10^{45}$ erg s$^{-1}$, we obtain
\begin{equation}
\label{nzero}
n_0 = 9.9 \times 10^9 (1 + \sigma)^{-1} \eta^{-1} L_{45} \Gamma_1^{-2} M_8^{-2} \Omega_{-1}^{-1} x^{-2} \ {\rm cm}^{-3} 
\end{equation}
and 
\begin{equation}
\label{Bzero}
B_0 =  1.1 \times 10^4 \sigma^{1/2} (1 + \sigma)^{-1/2} L_{45}^{1/2} \Gamma_1^{-1} M_8^{-1} \Omega_{-1}^{-1/2} x^{-1} \ {\rm G} . 
\end{equation} 
One may readily check whether the plasma frequency is much smaller than the electron cyclotron frequency. Under mean jet conditions,
\begin{equation}
\label{plasma}
\left({\nu_p \over \nu_c}\right)_0 = 3.2 \times 10^{-3}{n_0^{1/2} \over B_0} \sim 2.8 \times 10^{-2} \sigma^{-1/2}\eta^{-1/2} .  
\end{equation}
Note that this ratio does not depend explicitly on $r$. The ratio is even smaller, by a factor $\sim R^{-1}$, in masing regions where $B_m \sim R B_0$. Thus, the maser condition is satisfied at all radii provided that $\sigma\eta \gg 2.4\times 10^{-3} R^{-2}$.  (Note also that ion cyclotron damping need not quench the turbulence on scales larger than $r_0$, since the ion gyroradius is likely to be smaller than $r_0$ for the parameters under consideration, $r_i/ r_0 \sim (m_e T_i /m_i T_e)^{1/2} \sigma^{-1} \eta^{-1}$ [E. G. Zweibel, private communication].)

The observed frequency of the maser emission is related to the distance of the masers from the black hole and the characteristic mirror ratio $R$ through
\begin{equation}
\label{maserB}
B_0  \sim  36 \nu_{\rm GHz} R^{-1} \Gamma_1^{-1}   \ {\rm G} .
\end{equation}
Eliminating $B_0$ between equations (\ref{Bzero}) and (\ref{maserB}), we find that maser emission at a frequency $\nu$ is produced at
\begin{equation}
\label{xm}
r_{\rm obs} \sim  4.7 \times 10^{15} \sigma^{1/2} (1 + \sigma)^{-1/2} R \nu_{\rm GHz}^{-1} L_{45}^{1/2} \Omega_{-1}^{-1/2} \  {\rm cm}, 
\end{equation}
corresponding to $x_{\rm obs}\sim 10^3$ gravitational radii for $R \sim 5$ and with $M_8$ and all other parameters set equal to one.   

The typical jet density at $r_{\rm obs}$ is
\begin{equation}
\label{nzeroobs}
n_0 (r_{\rm obs}) = 9.4 \times 10^4 \sigma^{-1} \eta^{-1} \nu_{\rm GHz}^2 R^{-2} \Gamma_1^{-2} \ {\rm cm}^{-3} 
\end{equation}
and we can calculate the size, brightness temperature and power of a typical maser:
\begin{equation}
\label{pinchradobs}
r_m (r_{\rm obs}) \sim  6.5 \times 10^4 \sigma \eta \nu_{\rm GHz}^{-1} R^{2} \Gamma_1 \ {\rm cm} 
\end{equation}
\begin{equation}
\label{tbjet2}
T_B (r_{\rm obs})\sim 1.5 \times 10^{18}  \xi_{-2} \sigma^{-1}\eta^{-1} \nu_{\rm GHz}^{-1} R^{-2} \Gamma_1   \ {\rm K} 
\end{equation}
\begin{equation}
\label{tbjet3}
P_m (r_{\rm obs})\sim 2.9 \times 10^{17}  \xi_{-2} \sigma \eta R^{2}  \ {\rm erg \ s}^{-1}.
\end{equation}

In terms of the observationally inferred brightness temperature $T_{15} \equiv \langle T_B  \rangle_{\rm obs} / 10^{15}$ K, we find from eq.~(\ref{tbmean}) that the required covering factor of masers is
\begin{equation}
\label{cmjetobs}
C_m (r_{\rm obs}) \sim 6.7 \times 10^{-5} \xi_{-2}^{-1} \sigma \eta \nu_{\rm GHz} R^2 T_{15}  \Gamma_1^{-2} .  
\end{equation}
The masers need cover only a small fraction of the source on the sky, because each one has such a high brightness temperature. Another useful quantity is the volume filling factor.
Assuming that the depth of each maser along the line of sight is $\sim r_m$, the volume filling factor of masers is $f_m \sim (r_m / r) C_m$, where $r$ is the depth of the region containing the masers.  We presume this depth to be comparable to the distance of the masing region from the black hole at the base of the jet. Setting $r \sim r_{\rm obs}$, we have 
\begin{equation}
\label{fmjetobs}
f_m (r_{\rm obs})\sim 9.2 \times 10^{-16} \xi_{-2}^{-1} \sigma^{3/2} (1 + \sigma)^{1/2}\eta^2 \nu_{\rm GHz} R^3 T_{15} L_{45}^{-1/2} \Gamma_1^{-1} \Omega_{-1}^{1/2} . 
\end{equation}
The total number of maser sites required is 
\begin{equation} 
\label{fill1}
N_m (r_{\rm obs})\sim f_m \left({r_{\rm obs} \over r_m}\right)^3 \left({\Omega \over 4\pi}\right)\sim  
3.0 \times 10^{15} \xi_{-2}^{-1}(1 + \sigma)^{-1} \eta^{-1} \nu_{\rm GHz} T_{15} L_{45} \Gamma_1^{-4} \ . 
\end{equation}
Therefore, the total, broadband maser power emitted in the comoving frame of the jet (at $r \sim r_{\rm obs}$) is 
\begin{equation}
\label{powertot}
P_{\rm em} (r_{\rm obs})\sim N_m (r_{\rm obs})P_m (r_{\rm obs}) \sim  8.7 \times 10^{32} \sigma (1 + \sigma)^{-1}\nu_{\rm GHz} R^2 T_{15} L_{45} \Gamma_1^{-4}   \ {\rm erg \ s}^{-1},
\end{equation}
as one can verify directly from $\langle T_B  \rangle_{\rm obs}$ and $r_{\rm obs}$. The broadband ``isotropic power" inferred by an observer is boosted by a factor $\Gamma^4$.  These energy requirements are modest for the conditions thought to exist in blazar jets: for the model described here the volume-averaged emissivity of maser radiation at $r_{\rm obs}$ is only a fraction $\sim 2.5 \times 10^{-10} T_{15} R^2 \Gamma_1^{-2}$ of the comoving magnetic energy density passing through $r_{\rm obs}$ per dynamical time.

\subsection{Can cyclotron maser radiation escape from blazar jets?}

High--$T_B$ radiation at $\sim$ GHz frequencies, and cyclotron maser emission in particular, is subject to at least four processes that can inhibit the escape of radiation: 1) induced (stimulated) Compton scattering, 2) stimulated Raman scattering, 3) cyclotron absorption in the second and higher harmonics, and 4) synchrotron absorption by ultrarelativistic electrons. Synchrotron absorption is by far the most serious impediment, and will lead us to consider emission from a thin boundary layer.  

\subsubsection{Induced Compton and stimulated Raman scattering}

Once cyclotron maser radiation is produced it must traverse the gas in the non-masing regions along the line of sight, in order to escape to the observer.  Let us first consider the constraints imposed by induced Compton scattering and stimulated Raman scattering. We assume that we are looking down the jet, and that the main column density of intervening material is located close to the maser emission region, at $r_{\rm obs}$. Induced Compton scattering will limit the mean brightness temperature to a value that satisfies the condition 
\begin{equation}
\label{stimcomp1}
n_0 r \sigma_T {k \langle T_B  \rangle_{\rm obs} \over m_e c^2} \Gamma^{-2} \la 1,
\end{equation}
where $\sigma_T$ is the Thomson cross-section (Wilson 1982; de Kool \& Begelman 1989; Coppi, Blandford \& Rees 1993; Sincell \& Krolik 1994).  In this expression, one power of $\Gamma$ comes from the transformation of the observed brightness temperature to the comoving frame, and the other comes from the Lorentz contraction of the radial scale length $r$ (or, equivalently, from the Doppler-shifted scattering rate). Note that the relevant brightness temperature to use in evaluating the stimulated scattering rate is the angle-averaged mean brightness temperature in the comoving jet frame, rather than the higher $T_B$ associated with each maser beam.  This is because the scattering rate is proportional to the occupation number of available final states for the scattered photon, which includes an angular filling factor. Substituting $n_0$ from above we obtain an upper limit to the observed brightness temperature escaping from a given radius:
\begin{equation}
\label{tbstimcomp}
\langle T_B \rangle_{\rm obs }\la  6.0 \times 10^{12} (1+ \sigma) \eta L_{45}^{-1} \Gamma_1^4 M_8 \Omega_{-1}  x \ {\rm K} .
\end{equation}
Note the importance of relativistic beaming effects in facilitating the escape of radiation: a factor of 10 increase in $\Gamma$ increases the maximum brightness temperature by a factor of $10^4$.  If we set $x = x_{\rm obs}$ from eq.~(\ref{xm}), we obtain
\begin{equation}
\label{tbstimcomp1}
\langle T_B \rangle_{\rm obs }\la  1.9 \times 10^{15} \sigma^{1/2}(1+ \sigma)^{1/2}\eta \nu_{\rm GHz}^{-1} R L_{45}^{-1/2} \Gamma_1^4 \Omega_{-1}^{1/2}  \ {\rm K} .
\end{equation}
Thus, induced Compton scattering can pose a serious threat to the escape of cyclotron maser radiation in our picture, if $\langle T_B \rangle_{\rm obs}$ is much higher than $\sim 10^{15}$ K.  However, the difficulties are not insurmountable. As noted above, the effects of induced Compton scattering are greatly reduced if $\Gamma$ is even slightly higher; escape of radiation is also facilitated by larger $R$, larger ratio of Poynting flux to kinetic energy flux $\sigma$ (for fixed jet power), and larger jet solid angle $\Omega$.  Pair-dominated jets (i.e., with $\eta \gg 1$) have more trouble producing high $\langle T_B \rangle_{\rm obs }$, because of the larger electron and positron column densities required to carry a given $L_{KE}$.

Induced Compton scattering can also heat the electrons to temperatures well above the inverse Compton temperature associated with the total ambient radiation field (Levich \& Sunyaev 1970).  However, for the conditions we are considering one can show that ordinary cyclotron cooling exceeds induced Compton heating for all temperatures of interest.  Moreover, cyclotron cooling itself is not very important near $r_{\rm obs}$, since the cooling timescale exceeds the flow time.  Our model assumes that the mean temperature of electrons entering the pinch is mildly relativistic (to obtain reasonable maser efficiencies); such temperatures would have to be maintained by dissipative processes in the flow, such as reconnection or shocks.

We next consider stimulated Raman scattering, which was studied in detail by Levinson \& Blandford (1995).  As shown by their Fig.~6, Raman scattering becomes important relative to induced Compton scattering at low electron temperatures and high densities. (We note that Levinson \& Blandford considered Langmuir turbulence in a non-magnetized plasma, whereas we have the opposite situation, $\nu_c \gg \nu_p$.  Presumably, this difference between our case and theirs would tend to diminish the importance of Raman scattering.) At $x_{\rm obs}$, the mean electron density is $n_0(x_{\rm obs}) \sim 10^5 \sigma^{-1} \eta^{-1} \nu_{\rm GHz}^2 R^{-2} \Gamma_1^{-2}$ cm$^{-3}$. As shown by Blandford \& Levinson's Fig.~6, at such densities stimulated Raman scattering is completely unimportant compared to induced Compton scattering, for all values of $\langle T_B \rangle_{\rm obs}$ and $T_e$ under consideration.

\subsubsection{Local cyclotron absorption}

A major difficulty in applying the cyclotron maser to astrophysical radio sources is that emissions at the fundamental frequency are expected to be nearly entirely absorbed as they propagate through regions where they are at higher harmonics of the local electron gyrofrequency (Melrose \& Dulk 1982). There are several possible windows of escape: nearly parallel propagation (R-mode) where the optical depth is small, nearly perpendicular propagation with partial mode conversion, or mode conversion and other nonlinear processes in or near the absorption layer (Robinson 1989). We suggest that cyclotron absorption may be avoidable in current-carrying mirrors because the thickness of the second harmonic absorption layer is small.  A shell maser produces emission in the X-mode propagating perpendicular to ${\bf B}$. As the radiation moves away from the masing region, the magnetic field falls roughly as $B = B_m r_m/r$. Correspondingly, the maser emission travels through the second harmonic layer at $r = 2 r_m$. The damping rate ($\nu_d$) at this layer is unlikely reach as high as $\nu_p$, which in turn is much smaller than $\nu_c$. The thickness of the second harmonic damping layer can be estimated as $\sim r_m \nu_p/\nu_c$ so the travel time through the region is $\tau = r_m\nu_p/\nu_c c$ yielding an attenuation coefficient of $\sim r_m \nu_p^2/\nu_c c$.

From equation (\ref{pinchrad}), the size of the mirror can be expressed as $r_m \sim 0.1 c \nu_c/\nu_p^2$, so the attenuation coefficient is roughly 0.1. Thus, magnetic mirrors along current-carrying flux ropes not only form natural environments for the shell maser inversion process but also allow for the escape of the radiation because the regions of enhanced field are small and the field drops rapidly with transverse distance, resulting in a thin second harmonic absorbing layer.

\subsubsection{Synchrotron absorption}

If the magnetic field in the masing region is only a few times higher than the mean field outside, as we have argued, then there is no clearcut distinction between cyclotron absorption in the first few harmonics (where the absorbing electrons have $\gamma\ga 1$) and synchrotron absorption (where $\gamma \gg 1$) at radii $r \ga r_{\rm obs}$.  At radii $\gg r_{\rm obs}$ along the line of sight, where the magnetic field has dropped by a large factor, the dominant absorption will be squarely in the synchrotron regime, and we will make a relatively small error by using the synchrotron formulae everywhere. 

The flat radio spectra exhibited by many blazar jets are generally assumed to arise from self-absorbed synchrotron radiation.  Typical intrinsic brightness temperatures $\la 10^{11}$ K at a few GHz (Readhead 1994) indicate that the emission is produced where the mean magnetic field $B_0$ is considerably lower than in the masing regions, and therefore at distances $r \gg r_{\rm obs}$ from the black hole.  If this emission is truly self-absorbed and the emitting regions lie along the line of sight to the masers, then they will block the escape of the maser emission. 

Therefore, it should come as no surprise that GHz maser radiation can only escape from the interior of a jet with an anomalously low synchrotron emissivity (compared to the usual assumptions), corresponding to a very low ratio of relativistic electron energy density $U_{\rm rel}$ to magnetic energy density $U_B= B_0^2 /8\pi$.  Recall that standard models for self-absorbed jets (Readhead 1994) assume that $U_{\rm rel}$ and $U_B$ are close to equipartition.  If $U_{\rm rel} \ll U_B$, then the synchrotron emissivity saturates at a lower brightness temperature and the synchrotron photosphere lies closer to the black hole along the jet axis. (These arguments apply to the intrinsic brightness temperature; the perceived brightness temperature also depends on the Doppler factor, which could vary from jet to jet.) If the maser emission from the jet interior is blocked by synchrotron absorption, then we will see only the small fraction of emission produced outside the jet photosphere, presumably from a thin boundary layer along the wall of the jet at radii $\sim r_{\rm obs}$.

To estimate the energy density of relativistic electrons consistent with the escape of maser emission from the jet interior, let us suppose that the relativistic electrons have a power-law distribution in random Lorentz factor $n_{\rm rel} (\gamma)\propto \gamma^{-2}$ for $\gamma_{\rm min} \leq \gamma \leq \gamma_{\rm max}$.  If $\gamma_{\rm min}$ is small enough that the corresponding synchrotron critical frequency is below the observing frequency (as Doppler-shifted into the comoving frame), $\gamma_{\rm min} < \gamma_c (\nu, B_0) = 5 B_0^{-1/2} \nu_{\rm GHz}^{1/2} \Gamma_1^{-1/2}$ (corresponding to $\gamma_{\rm min }\la r/ r_{\rm obs}$), then the synchrotron absorption coefficient at the observing frequency $\nu_{\rm GHz}$ GHz is given by (Rybicki \& Lightman 1979)
\begin{equation}
\label{synchabs}
\alpha_\nu \sim 2.2\times 10^{-9} {B_0^4 \over \ln (\gamma_{\rm max} / \gamma_{\rm min}) } {U_{\rm rel} \over U_B } \Gamma_1^3 \nu_{\rm GHz}^{-3}   \ {\rm cm}^{-1}. 
\end{equation}
In the magnetically dominated jet environment we might reasonably expect the energy density of relativistic electrons not to exceed the thermal energy density, $U_{\rm th}\sim 3 n_0 k T_e/2$. Normalizing $U_{\rm rel}$ to $U_{\rm th}$ instead of $U_B$ we obtain the optical depth along the line of sight to radius $r > r_{\rm obs}$, 
\begin{equation}
\label{synchtau}
\tau_{\rm synch} (r) \sim 7.8\times 10^{9} {\sigma^{-1/2}(1 + \sigma)^{-1/2} \over \ln (\gamma_{\rm max} / \gamma_{\rm min}) } \eta^{-1}{U_{\rm rel} \over U_{\rm th} }{kT_e \over 100 \ {\rm keV}} R^{-3} \Gamma_1^{-1}  L_{45}^{1/2} \Omega_{-1}^{-1/2} \left({r \over r_{\rm obs}}  \right)^{-3} .
\end{equation}

If $\gamma_{\rm min} > \gamma_c$, both the absorption coefficient and the optical depth are lowered by a factor $(\gamma_{\rm min}/ \gamma_c)^{-8/3}$, implying that even a moderate increase in $\gamma_{\rm min}$ can greatly suppress the absorption of maser radiation.  Nevertheless, it is clear that synchrotron absorption provides by far the most serious impediment to the escape of maser radiation.  To have any hope of observing high brightness temperature emission, we must consider rather ``optimistic" values of the main parameters.  One might reasonably assume $R \sim 5$ (although higher values are not excluded), $\ln (\gamma_{\rm max} / \gamma_{\rm min}) \sim 10$, and $\eta \sim 1$ (electron-ion plasma).  We shall take $kT_e \sim 100$ keV and --- more speculatively --- suppose that  $U_{\rm rel}$ could be as low as $\la 10^{-3 }U_{\rm th}$, which is consistent with a Maxwellian tail on the thermal electron distribution (although Compton cooling on diffuse ambient radiation could deplete such a tail).  The optical depth is reduced to 
\begin{equation}
\label{synchtau2}
\tau_{\rm synch} (r_{\rm obs}) \sim {6.2\times 10^{3} \over  \sigma^{1/2}(1 + \sigma)^{1/2} } {U_{\rm rel} \over 10^{-3}U_{\rm th} } \Gamma_1^{-1} L_{45}^{1/2} \Omega_{-1}^{-1/2}  
(1 +  \gamma_{\rm min}^{8/3} )^{-1},
\end{equation}
Note that $\tau_{\rm synch} (r_{\rm obs})$ is independent of frequency, since the decrease of column density with radius is offset by the decreasing frequency of the emission.   

There are combinations of parameters --- large values of $\sigma$, $\Gamma$ and $\gamma_{\rm min}$; or smaller values of $U_{\rm rel}$ --- that lead to $\tau_{\rm synch} < 1$, thus allowing the maser emission to escape from the interior of the jet. None of these quantities (with the possible exception of $\Gamma$) is even moderately well-constrained, especially at radii $\sim r_{\rm obs}$ that are smaller than the synchrotron emission radius.  Moreover, bright maser emission may still be observable even if $\tau_{\rm synch} > 1$.  For example, the maser emission could escape from a thin boundary layer along the wall of the jet.  The fraction of the jet volume involved in producing observable maser radiation at $r$ would then be $\sim 1/ \tau_{\rm synch} (r)$; the filling factor of maser sites in the boundary layer would have to be correspondingly higher.  Indeed, preferential formation of maser-producing mirrors near the jet boundary is not implausible, since strong shear, tending to stretch and compress the field, is likely to be especially prevalent near the walls of the jet.  An attractive feature of the boundary layer model is that shear at the interface with the ambient medium could drive supermagnetosonic turbulence, leading to localized magnetic field strengths far higher than the mean jet field.  This would allow the observed maser radiation to be produced at larger radii, facilitating its escape. 

While it would certainly help for maser emissivity to be enhanced close to the walls of the jet, compared to the interior, it is by no means necessary. This is because the maser mechanism is so efficient: for maser emissivity spread randomly through the jet, the ratio of volume-averaged maser emissivity to magnetic energy supply need only be $\sim 10^{-8}T_{15}\Gamma_1^{-2}$ to provide the observed brightness temperature.  If we only see the masers in a boundary layer, the required emissivity is increased by a factor $\tau_{\rm synch} (r_{\rm obs})$ but is still quite small.  We note that these estimates do not take into account the possibility of synchrotron absorption and induced Compton scattering in the ambient medium that surrounds the jet, the nature of which is very uncertain. 

\section{Discussion}  

We have outlined a set of physical processes by which high-brightness temperature electron cyclotron maser emission can be produced in blazar jets, as well as other tenuous, highly magnetized astrophysical media.  The population inversion leading to maser emission occurs in transient, small-scale, current-carrying magnetic mirrors with mirror ratios $\ga 3$. The latter can arise from hydromagnetic instabilities, shocks, and/or turbulence driven, for example, by shear near the interface between the jet and the ambient medium.  The mechanism we have described is an example of a ``shell maser", in which the population inversion results mainly from electron acceleration in a parallel electric field; the more commonly discussed ``loss cone" effects are secondary. As a precondition for efficient maser emissivity, a jet should have a ratio of Poynting flux to kinetic energy flux that is not too small, so that the electron cyclotron frequency exceeds the electron plasma frequency by a large factor.  Another condition for high maser efficiency is that the thermal electrons carrying the current into the magnetic mirrors be hot, with $kT_e \sim 100$ KeV.  The mechanism works much better in a jet composed of electron-ion plasma than in a pair-dominated jet, since the latter will have a higher ratio of plasma frequency to cyclotron frequency and a larger induced Compton scattering depth. 

We are unable to estimate the volume fraction of a turbulent jet that will have properties conducive to maser emission, but we show that GHz brightness temperatures $\sim 10^{15}$ K can arise from systems with an extremely small filling factor of such sites.  Moreover, the fraction of internal (magnetic) energy that needs to be converted to maser radiation per flow time is extremely small.  These conclusions are encouraging given the severe constraints on the escape of maser radiation from the jet, once it is produced.  Induced Compton scattering places upper limits on the brightness temperature of the escaping radiation; however, the main limitation comes from synchrotron absorption by relativistic electrons within the jet. For all but the most extreme parameters the jet will be very optically thick to maser radiation, but we argue that a thin boundary layer along the wall of the jet could produce high brightness temperatures without excessive demands on the energetics or filling factor of masing regions.  We note that even the boundary layer model requires relativistic electron energy densities far below equipartition --- this could be due to weak particle acceleration in these inner jet regions, where little of the magnetic energy has been dissipated, possibly combined with the effects of rapid synchrotron or inverse Compton cooling. 

Our model is parameterized in terms of maser emission at a few GHz, but the process could clearly operate at other frequencies as well. According to our jet model, the radius at which maser radiation of a given frequency originates [$r_{\rm obs} (\nu)$] scales as $\nu^{-1}$.  Thus, maser emission at frequencies of up to $\sim 100$ GHz could be produced near the bases of blazar jets.  The energy efficiency of maser emission (defined at the end of \S 3.1) is independent of $\nu$, as is the characteristic synchrotron optical depth.  However, the maximum brightness temperature permitted by induced Compton scattering scales as $\nu^{-1}$. 

The relations derived in \S 3 can be scaled to relativistic jets from stellar-mass systems, such as microquasars.  If we assume that the jet power, $L$, scales linearly with the mass of the black hole, then the frequency of maser radiation produced at a given dimensionless radius, $x_{\rm obs}$ (in units of the gravitational radius $GM/c^2$), scales as $\nu \propto M^{-1/2}$.  Therefore, microquasars (and other types of X-ray binary) could emit cyclotron maser emission in the infrared or optical band, from the base of the jet or a corona above the accretion disk.  The characteristic synchrotron optical depth scales as $M^{1/2}$, independent of $\nu$; therefore, typical values of $\tau_{\rm synch}$ should be 3--4 orders of magnitude lower than those estimated for blazars.  This greatly relaxes the constraints on boundary layer emission, and raises the possibility that microquasar jets may be optically thin to maser radiation from their interiors.  The induced Compton limit on the brightness temperature emitted at a given $x_{\rm obs}$ is independent of mass, but at a fixed observing frequency the brightness temperature limit scales as $M^{-1/2}$.  Thus, high brightness temperature maser emission has an easier time escaping from lower mass systems.
 
The principal diagnostics of planetary and stellar cyclotron maser emission --- high circular polarization and extreme variability --- are probably not useful for analyzing cyclotron maser emission from jets.  The enormous number of independent maser sites, spread over a region vastly larger than the size of each maser, would presumably wash out such signatures.  Clues to a coherent emission mechanism could be obtained by brightness temperature measurements as a function of frequency as well as evidence for an unusual source geometry, such as a narrow boundary layer.  Although the production and escape of cyclotron maser radiation from relativistic jets is by no means inevitable, the search for coherent emission remains exceedingly worthwhile because of the constraints it would impose on the structure and composition of these energetic flows.

\begin{acknowledgements} 
We are indebted to Ellen Zweibel, Roger Blandford and Paolo Coppi for their insightful comments and criticisms, and to Jean-Pierre Macquart for bringing the latest scintillation measurements to our attention. MCB acknowledges support from National Science Foundation grant AST-0307502.   

\end{acknowledgements}

\begin{figure} 
\plotone{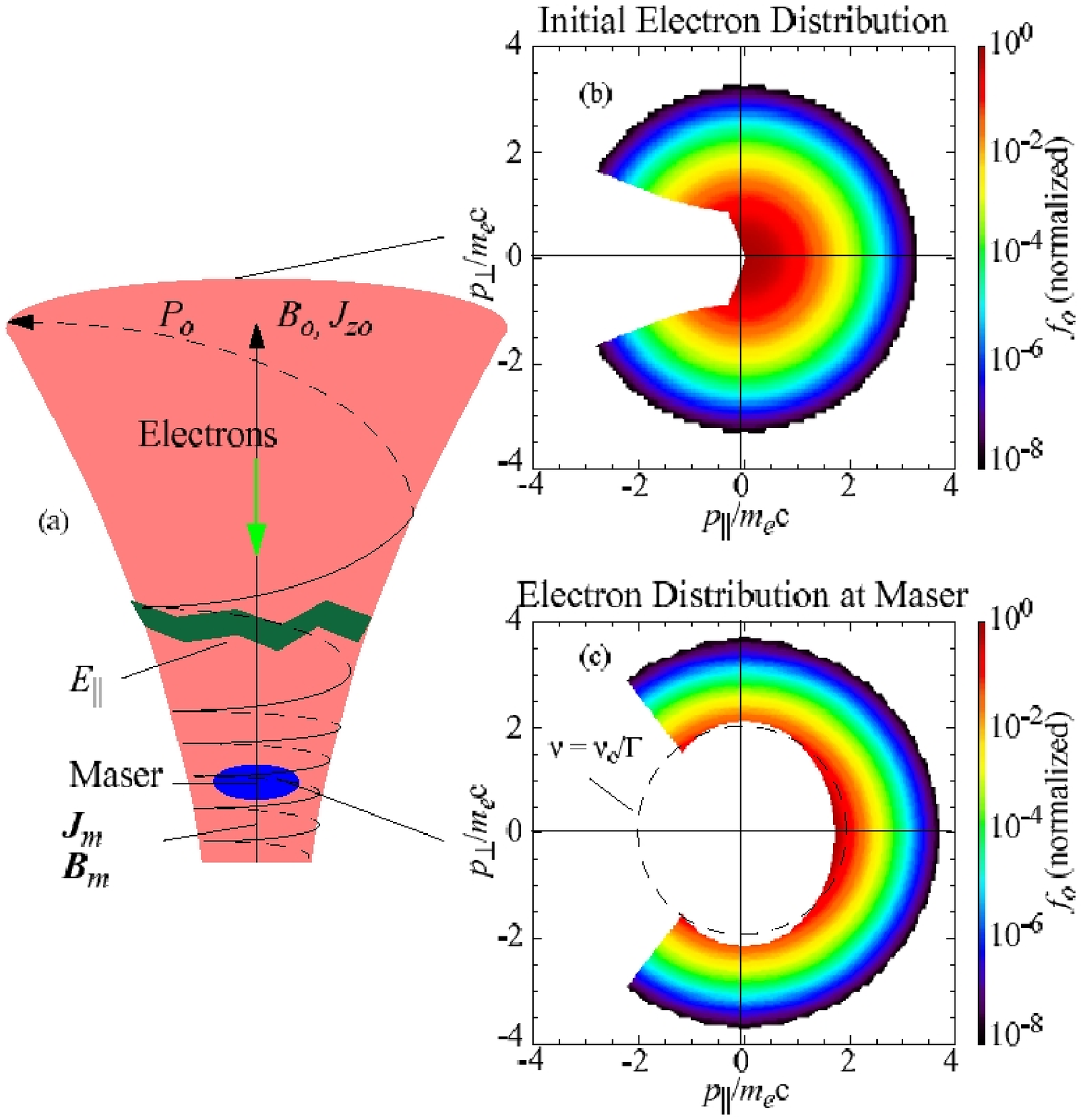}
\caption{  Fig. 1. (a) Cartoon of a current-carrying magnetic mirror on a quasi-force-free flux rope. The nearly field-aligned current increases both the toroidal and axial components of the magnetic field. Electrons carry the current of the flux rope from top to bottom. A parallel electric field, as in the aurorae on Earth and Jupiter, is required to maintain the electron flow through the mirror region. The combination of parallel potential and magnetic mirror evolves the initial electron distribution (b) into a horseshoe-shaped distribution (c) when viewed in 2-D (a shell when viewed in 3-D). In these plots we use a mirror ratio $R=5$, a current $|J_{zm}| = 30 mA/m^2$ ($|J_{z0}| = 6 mA/m^2$), and a 500 keV parallel potential which is consistent with the current for an electron temperature of $\sim 100$ keV and density of 100 cm$^{-3}$. The adiabatic, static Vlasov code used for these calculations is described by Ergun et al. (2000b). 
\label{fig1} }
\end{figure}

\end{document}